\documentclass[prb,twocolumn,superscriptaddress]{revtex4}

\usepackage{graphicx}
\usepackage{dcolumn}
\usepackage{amsmath}
\usepackage{color}

\begin{document}

\title{Electronic structure and magnetic correlations in the epitaxially strained bilayer nickelate La$_3$Ni$_2$O$_{7}$}

\author{I. V. Leonov}
\affiliation{M. N. Mikheev Institute of Metal Physics, Russian Academy of Sciences, 620108 Yekaterinburg, Russia}
\affiliation{Institute of Physics and Technology, Ural Federal University, 620002 Yekaterinburg, Russia}

\begin{abstract}

Using the DFT+dynamical mean-field theory method we study the effects of electron-electron correlations and epitaxial strain of the crystal structure on the normal-state electronic structure, quasiparticle band renormalizations, Fermi surface, and magnetic correlations of the bilayer Ruddlesden-Popper nickelate La$_3$Ni$_2$O$_7$ (LNO). Our results exhibit a remarkable orbital-selective renormalization and strong incoherence of the Ni $3d$ bands, pointing to the proximity of the Ni $x^2-y^2$ and $3z^2-r^2$ states to orbital-selective localization. The electronic properties of LNO show a high sensitivity to the in-plane strain. We note that both a tensile and a moderate compressive strain (up to about $-2$\%) yield a significant enhancement of magnetic correlations compared to the unstrained LNO. Under a large compressive strain of about $-4$\%, we observe a Lifshitz transition characterized by the disappearance of the $\gamma$ Fermi surface sheet, which is associated with a nearly fully occupied, shallow flat-band of the bonding Ni $3z^2-r^2$ orbital character. As a result, we observe a sharp decrease of magnetic correlations, implying suppression of superconductivity. Overall, our results support the picture of spin- and change-density-wave stripe instability driven by the Fermi surface nesting in LNO. Our results suggest that both pressure and strain can effectively tune (suppress or enhance) spin-change-density-wave ordering, giving rise to enhanced spin fluctuations. 

\end{abstract}

\maketitle

\section{Introduction}

The recent discovery of unconventional superconductivity in the highly pressurized bilayer and trilayer Ruddlesden-Popper nickelates Ln$_{n+1}$Ni$_n$O$_{3n+1}$ (LNO) \cite{Sun_2023,Hou_2023,Wang_2024a,YZhang_2024,Wang_2024b,Li_2025,Qiu_2025,
MWang_2024,Zhu_2024,Li_2024,Zhang_2025,MZhang_2025,
Pei_2026},
with $\mathrm{Ln}=\mathrm{La}$, Pr, and Sm, and $n=2$ and $n=3$, respectively, has generated great research interest to this novel class of superconducting compounds \cite{Dong_2024,Yang_2024a,Liu_2024,Xie_2024,MZhang_2024,Liu_2025,Zhang_2023a,
Shilenko_2023,Lechermann_2023,Christiansson_2023,Liao_2023,Shen_2023b,Ryee_2024,
Craco_2024,Cao_2024,YYang_2023,Qin_2023,Leonov_2026,
Wang_2024,Leonov_2024a,LaBollita_2024a,Huang_2024,Chen_2024b,Tian_2024b,
ZhangLin_2024b,Yang_2024b}. It has been demonstrated that superconductivity sets below a high critical temperature of about 80-90~K in the bulk bilayer and $\sim$30-40~K in the bulk trilayer LNOs (with a partial substitution of La with Pr and Sm) under pressure above 15~GPa. It appears near a pressure-driven structural phase transition to the high-symmetry (orthorhombic or tetragonal) phase, which is characterized by a change of the Ni-O-Ni bond angle from about 166$^\circ$ to 180$^\circ$ (along the $c$ axis), with the absence of tilting of NiO$_6$ octahedra in the superconducting phase \cite{Sun_2023,Hou_2023,Wang_2024a,YZhang_2024,Wang_2024b,Li_2025,Qiu_2025,
MWang_2024,Zhu_2024,Li_2024,Zhang_2025,MZhang_2025,
Pei_2026}. 

In these materials a nominal electronic configuration is of $3d^{7.5}$ (Ni$^{2.5+}$) for the bilayer and $3d^{7.33}$ (Ni$^{2.67+}$) for the trilayer systems, characterized by the partially occupied Ni $x^2-y^2$ and $3z^2-r^2$ bands near the Fermi level \cite{Zhang_2023a,Shilenko_2023,Lechermann_2023,Christiansson_2023,Liao_2023,
Shen_2023b,Ryee_2024,Craco_2024,Cao_2024,YYang_2023,Qin_2023,
Leonov_2026,Wang_2024,Leonov_2024a,LaBollita_2024a,Huang_2024,Chen_2024b,
Tian_2024b,ZhangLin_2024b,Yang_2024b}. In fact, the low-energy electronic states of these materials (in the paramagnetic metallic state) exhibit multi-orbital Fermi surface with distinct electron and hole pockets \cite{Li_2017,B.Y.Wang_2025, P.Li_2025,Au-Yeung_2026,W.Sun_2026,J.Shen_2026}. Moreover, these materials are characterized by strong covalency between the Ni $3d$ and O $2p$ orbital states, implying the importance of small (or negative) charge-transfer energy effects \cite{Zaanen_1985,Shilenko_2023,Lechermann_2023,Zhang_2026,Chen_2026,Li_2026}. Superconductivity in the high-pressure phase is found to border with a strange (bad) metal phase in the normal state above $T_c$, whereas the electronic structure exhibits high sensitivity to crystalline quality and chemical composition (e.g., oxygen nonstoihiometry and doping) \cite{Y.Miao_2025,M.Shi_2025,M.Shi_2025b}. All together, this suggests the importance of the effects of strong orbital-selective correlations in the partially filled Ni $3d$ shell of LNO. 

Moreover, recent experiments, such as transport, resonant x-ray scattering, inelastic neutron diffraction, muon spin relaxation, nuclear magnetic and quadrupole resonance,  and optical spectroscopy measurements, show anomalous behavior of the properties of the bilayer and trilayer LNOs near $\sim$150~K (at ambient pressure) \cite{Chen_2024a,Kakoi_2024,Agrestini_2024,Dan_2024,Meng_2024,Khasanov_2024,
Cao_2025, LiGong_2025,Dou_2026}. It has been attributed to the emergence of the complex double spin-charge-density wave stripe ordering characterized by diagonal hole stripes (in the NiO$_2$ plane) oriented at 45$^\circ$ to the Ni-O bond with a propagating vector $\mathrm{\bf q}=(\frac{1}{4},\frac{1}{4})$, with zigzag ferromagnetic chains alternating in the $ab$ plane \cite{Leonov_2025,LaBollita_2024b,BZhang_2024,Ni_2024,Tian_2025}. Under pressure, the stripe ordering is found to collapse at a critical pressure threshold coinciding with the structural phase transition into the high-$T_c$ superconducting phase \cite{Sun_2023,Hou_2023,Wang_2024a,YZhang_2024,Wang_2024b,Li_2025,Qiu_2025,
MWang_2024,Zhu_2024,Li_2024,Zhang_2025,MZhang_2025,
Pei_2026}. This implies that suppression of the spin-charge-density wave state under pressure paves the way for the emergence of superconductivity in these systems. Note that in-plane spin-charge-density wave stripe orderings (with different periodicity) accompanied by breathing-mode distortions of the crystal structure have been previously discussed in the context of layered nickelates \cite{Yamada_1992,Fabbris_2017,Petsch_2023,K.Slobodchikov_2022}. This suggests that spin- and charge-density wave stripe states, as well as their fluctuations are crucial for the understanding of the low-energy properties of LNOs.

Very recently, superconductivity with a high $T_c$ exceeding 40~K has been experimentally observed in the bilayer nickelate thin films at ambient pressure \cite{Ko_2025,Zhou_2025,Y.Liu_2025}. It was shown that superconductivity can be stabilized in the high-quality bilayer LNO thin films deposited on the LaSrAlO$_4$ substrate which maintain a tetragonal phase with an uniform epitaxial compressive strain $\sim$$-2$\% in the NiO$_2$ planes (relative to the bulk) \cite{Ko_2025,Zhou_2025,Y.Liu_2025}. This breakthrough opened a new avenue for employing various experimental probes to study the electronic structure, magnetism, and superconductivity in these systems \cite{Y.Zhang_2026}. In particular, recent angle-resolved photoemission spectroscopy measurements on the bilayer LNO thin films \cite{B.Y.Wang_2025, P.Li_2025,Au-Yeung_2026,W.Sun_2026,J.Shen_2026} show the presence of the $\alpha$ and $\beta$ Fermi surface sheets, originating from the strongly hybridized Ni $x^2-y^2$ and $3z^2-r^2$ orbitals, in close similarity to the bulk bilayer LNO \cite{Shilenko_2023,Lechermann_2023,Leonov_2026}. However, the existence of the $\gamma$ Fermi surface pocket, previously extensively discussed in the context of the electronic structure of the bulk bilayer LNO, remains a subject of debate \cite{B.Y.Wang_2025, P.Li_2025,Au-Yeung_2026,W.Sun_2026,J.Shen_2026}. The latter is associated with a nearly fully occupied, shallow flat-band of the bonding Ni $3z^2-r^2$ orbital character, which seems play a decisive role for the appearance of superconductivity \cite{Leonov_2026}. Moreover, recent experiments also show that $T_c$ can be further enhanced by applying hydrostatic pressure to the compressively strained  bilayer nickelate thin films \cite{Q.Li_2026}.

While experimental and theoretical analysis point to anisotropic $s$-wave type superconductivity (possibly competing with the $d$-wave component in the bulk materials), the microscopic origins of this anomalous behavior remain a matter of debate \cite{Zhang_2023c,Liu_2023b,Yang_2023,Heier_2024,Lu_2024,Fan_2024,Sakakibara_2023a,
Tian_2024a}. In fact, despite recent active research \cite{Y.F.Zhao_2025,B.Geisler_2025,B.Geisler_2026a,B.Geisler_2026b,
Y.Zhang_2026b,L.Bleys_2025,X.W.Yi_2025,Shao_2025a,Shao_2025b,
W.Qiu_2026,Y.Zhong_2026,C.Le_2025,Ushio_2026,Y.H.Cao_2026}, the effects of epitaxial strain and electron-electron correlations on the electronic structure, quasiparticle band renormalization, and magnetic correlations in the bilayer LNO remain largely unexplored. We address this topic in our present study. 


\section{Results and discussion}

In this paper, we report a theoretical study of the effects of electron-electron correlations and epitaxial strain on the normal-state electronic structure, quasiparticle renormalizations, Fermi surface, and magnetic correlations in the bilayer Ruddlesden-Popper LNO. In our work, using a fully charge self-consistent implementation of the DFT+dynamical mean-field theory method \cite{Georges_1996,Kotliar_2006,Leonov_2020a,Leonov_2024b} ($\mathrm{DFT}+\mathrm{DMFT}$) we explore the normal-state electronic structure properties of the paramagnetic (PM) phase of LNO. We examine a biaxially strained crystal structure of LNO with the in-plane strain tuned from $-4$\% (compressive) to +4\% (tensile), with a focus to the strain-driven changes in the electronic structure, charge transfer effects between the Ni $x^2-y^2$ and $3z^2-r^2$ orbital states, orbital-selective quasiparticle  renormalizations, Fermi surface, and magnetic correlations of PM LNO. 

In our calculations, we use the orthorhombic $Fmmm$ crystal structure with the lattice constants and atomic coordinates determined from structural optimization of LNO within nonmagnetic DFT. In DFT we employ generalized gradient approximation with the Perdew-Burke-Ernzerhof (PBE) exchange functional as implemented in the 
Quantum ESPRESSO package\cite{Giannozzi_2009,Giannozzi_2017,DalCorso_2014}. In the calculations, the in-plane ($ab$ plane) lattice vectors were fixed, whereas the $c$-lattice parameter and the atomic coordinates were fully relaxed (within nonmagnetic DFT). Note that the unstrained structure is not the same as the strain-free bulk structure of LNO, as the latter is not compatible with the epitaxial constraints. In fact, at ambient pressure PM LNO crystallizes in the orthorhombic $Amam$ crystal structure, which is characterized by tilting of the Ni-O-Ni bond angles along the $c$ axis \cite{Sun_2023,Hou_2023,Wang_2024a,YZhang_2024,Wang_2024b,Li_2025,Qiu_2025,
MWang_2024,Zhu_2024,Li_2024,Zhang_2025,MZhang_2025,
Pei_2026}.

Our results for the structural optimization are in overall good agreement with available experimental data \cite{Ko_2025,Zhou_2025,Y.Liu_2025}. We find that applying a biaxial compressive strain of $-2$\% results in a small elongation of the unit cell by about 1.7\% (along the $c$ axis), with the $c$ axis lattice constant of about 20.53 \AA, compared to the unstrained system of 20.17~\AA. The unit cell volume is found to slightly decrease by about 2\%. The calculated in-plane lattice constant is about 3.747~\AA\ (compared to 3.823 \AA\ for the unstrained $Fmmm$ lattice). This result agrees well with the recent x-ray diffraction refinement of the crystal structure of the compressively strained thin films of bilayer LNO grown on the LaSrAlO$_4$ (001) substrate \cite{Ko_2025,Zhou_2025,Y.Liu_2025}. In contrast, an applying of a biaxial tensile strain of 2\% leads to a decrease of the $c$-axis lattice constant by $\sim$1.4\% to about 19.89~\AA, with the in-plane lattice constants of 3.67 \AA. In addition, the unit cell volume slightly increases by about 2.5\%. The obtained data on the crystal structure are consistent with the bilayer LNO thin films grown on the Ba$_{1-x}$Sr$_x$O (with $x\sim 0.25$) or SrO-SrTiO$_3$ (001) substrate \cite{L.Liu_2025}.

We use the $\mathrm{DFT}+\mathrm{DMFT}$ method \cite{Georges_1996,Kotliar_2006} in order to treat many-body correlation effects in the partially filled Ni $3d$ states (the La $5d$
states are more extended, located well above the Fermi level, and therefore can be considered as uncorrelated). We utilize a fully charged self-consistent $\mathrm{DFT}+\mathrm{DMFT}$ scheme implemented with plane-wave pseudopotentials to compute the orbital-dependent and {\bf k}-resolved spectral functions of PM LNO \cite{Leonov_2020a,Leonov_2024b}. In our calculations we neglect  the possible occurrence of spin and charge density wave states in LNO. For the low-energy states we construct a basis set of atomic-centered Wannier functions for the Ni $3d$, La $5d$, and O $2p$ valence states using the energy window spanned by these bands \cite{Anisimov_2005,Marzari_2012}. This allows us to treat the electron-electron interactions in the partially filled Ni $3d$ shell, complicated by a charge transfer between the Ni $3d$, O $2p$, and La $5d$ states. The O $2p$ and La $5d$ states are treated as uncorrelated on the DFT level within the fully charge self-consistent $\mathrm{DFT}+\mathrm{DMFT}$ method.

In $\mathrm{DFT}+\mathrm{DMFT}$ we employ the continuous-time hybridization expansion quantum Monte Carlo algorithm (CT-QMC) \cite{Gull_2011} with improved estimators for the self-energy \cite{Boehnke_2011,Hafermann_2012} in order to describe a realistic many-body problem associated with the strongly correlated Ni $3d$ electrons. In agreement with previous applications of $\mathrm{DFT}+\mathrm{DMFT}$ to study the Ruddlesden-Popper nickelates \cite{Lechermann_2023,Shilenko_2023,Leonov_2024a,
Leonov_2026}, we use the Hubbard $U = 6$~eV and Hund's exchange coupling $J = 0.95$ eV, together with the fully localized double-counting correction (evaluated from the self-consistently determined local occupations). In our calculations we neglect the spin-orbit coupling, which is assumed to be negligible for the Ni $e_g$ orbitals. In order to compute the {\bf k}-resolved spectra, quasiparticle band renormalizations, and correlated Fermi surfaces we perform analytic continuation of the self-energy results using Pad\'e approximants. Our $\mathrm{DFT}+\mathrm{DMFT}$ calculations are performed for the normal state of PM LNO at a temperature $T = 290$~K.

\begin{figure}
\centerline{\includegraphics[width=0.5\textwidth,clip=true]{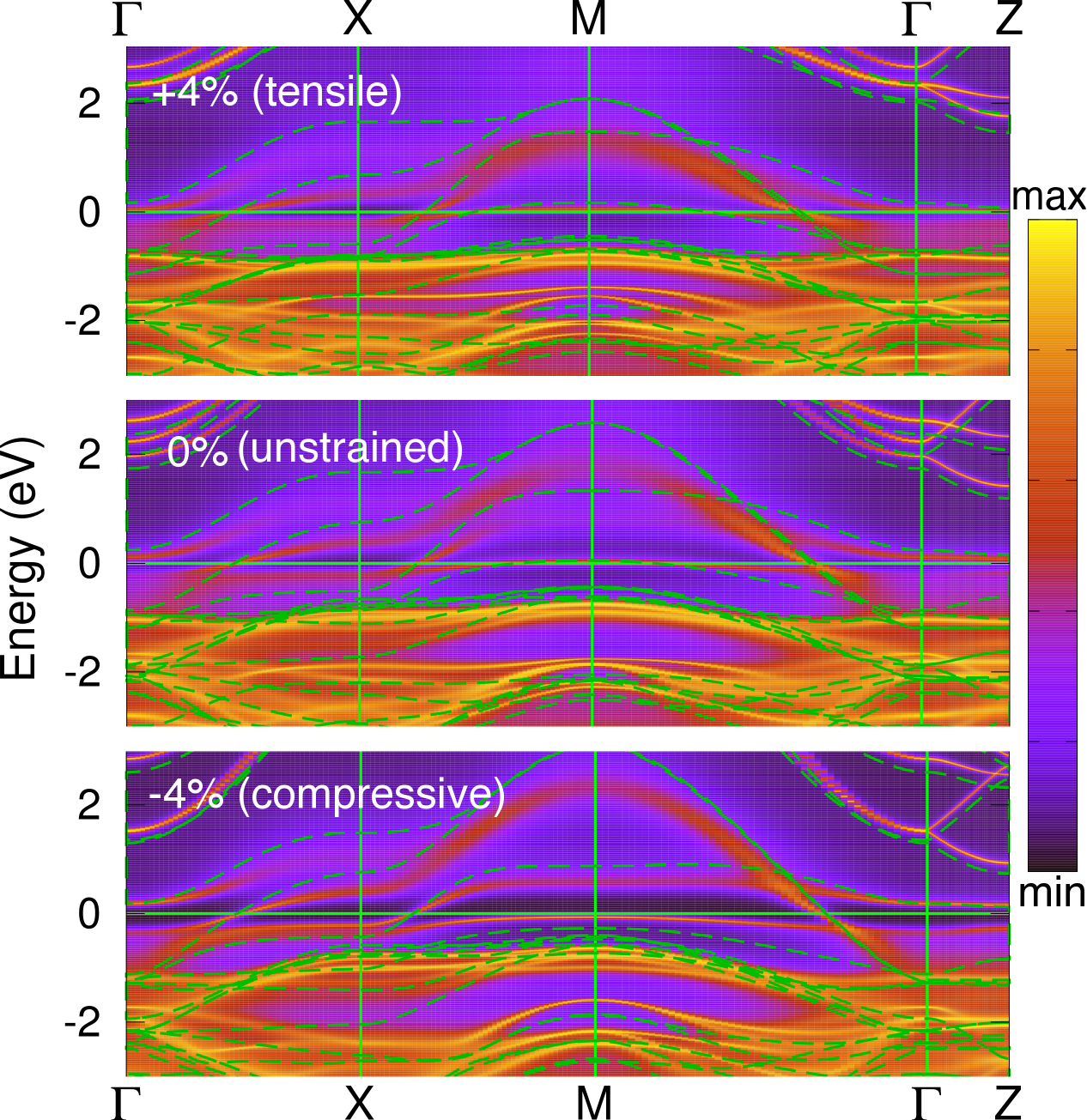}}
\caption{{\bf k}-resolved total spectral functions $A({\bf k},\omega)$ of PM LNO under different strains calculated by $\mathrm{DFT}+\mathrm{DMFT}$ with $U=6$ eV and $J=0.95$ eV at $T = 290$~K. Our results are compared with those obtained within the nonmagnetic DFT calculations (shown by dashed green lines).
}
\label{Fig_1}
\end{figure}

\begin{figure}
\centerline{\includegraphics[width=0.4\textwidth,clip=true]{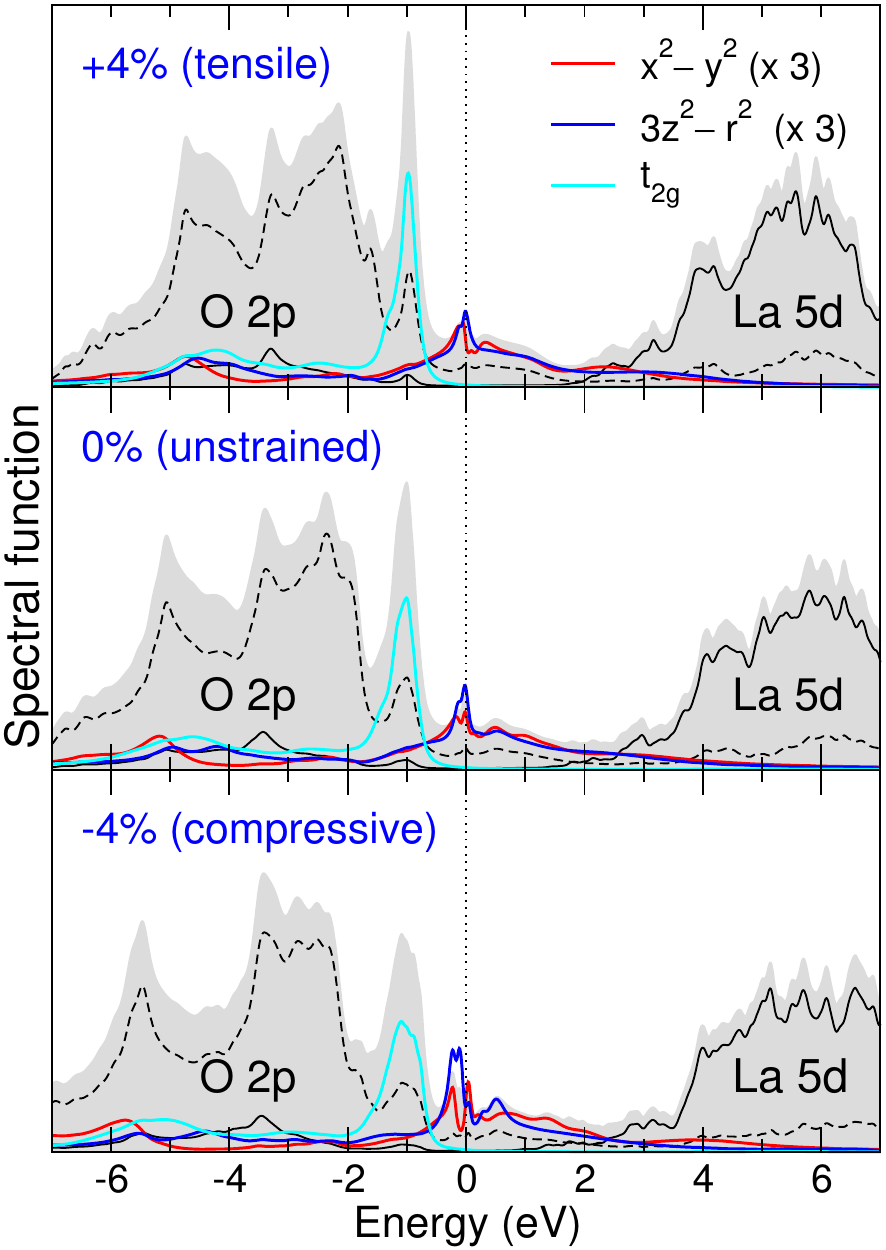}}
\caption{
Orbital-dependent spectral functions for PM LNO for different in-plain lattice strains as obtained by $\mathrm{DFT}+\mathrm{DMFT}$ at $T = 290$~K. The Ni $3d$, La $5d$ and O $2p$ orbital contributions are shown. The partial Ni $x^2-y^2$, and $3z^2-r^2$ orbital states are magnified by three times for better readability.
}
\label{Fig_2}
\end{figure}

We start by computing the electronic structure and analysing the orbital-selective behavior of the Ni $3d$ states.
In Fig.~\ref{Fig_1} we display our results for the {\bf k}-resolved spectral functions obtained using $\mathrm{DFT}+\mathrm{DMFT}$ for PM LNO under the 
tensile (+4\%), unstrained (0\%), and compressive lattice strain ($-4$\%). For comparison we show the nonmagnetic 
DFT band structure results. The corresponding {\bf k}-integrated orbital-dependent spectral functions are shown in Fig.~\ref{Fig_2}. In agreement with previous studies, the electronic states near the Fermi level are primarily derived from the Ni $x^2-y^2$ and $3z^2-r^2$ orbitals. Our results exhibit a typical for the bilayer LNO bonding-antibonding splitting of the partially occupied Ni $e_g$ bands near the Fermi level, which is associated with a strong interlayer coupling of the Ni $3z^2-r^2$ orbitals. The bonding-antibonding splitting of the Ni $3z^2-r^2$ orbitals in LNO is about 1.16 eV (shown Table~\ref{tab_1}), close to that value in the bulk high-pressure phase of LNO ($\sim$1.3~eV). This suggests that the interlayer hopping between the Ni $3z^2-r^2$ orbitals is still relatively large, about $-579$~meV, sufficient for a strong inter-layer superexchange coupling that can be effectively tuned by the strain effects. For comparison, the Ni $3z^2-r^2$ inter-layer hopping for the bulk high-pressure phase of LNO at $\sim$30 GPa is about $-630$~meV. The fully occupied Ni $t_{2g}$ states appear below the Fermi level, between $-2$ and $-1$ eV. The O $2p$ orbital states strongly hybridize with the Ni $3d$ and La $5d$ states, appear about $4$ eV below $E_F$. The rare-earth element La $5d$ orbitals are empty and are located above about 2 eV. 

\begin{figure}
\centerline{\includegraphics[width=0.5\textwidth,clip=true]{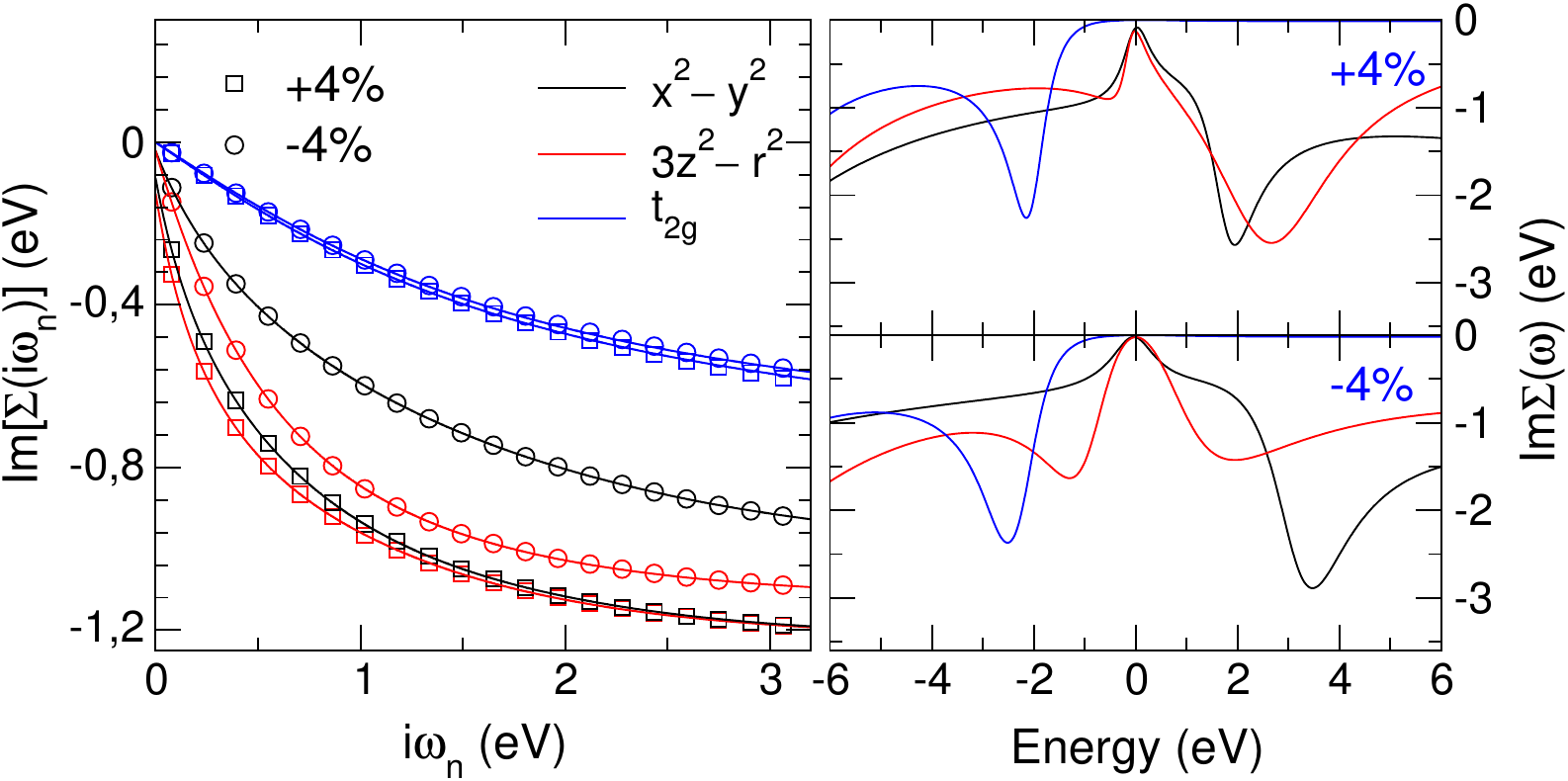}}
\caption{
Orbital-dependent contributions to the imaginary part of the Ni $3d$ self-energy on the Matsubara axis, $\mathrm{Im}[\Sigma(i\omega_n)]$ (left, symbols), and analytically continued on the real energy axis, $\mathrm{Im}[\Sigma(\omega)]$ (right), for PM LNO under the 4\% (tensile) and $-4$\% (compressive) biaxial lattice strain at $T = 290$~K. The Pad\'e interpolation results for $\mathrm{Im}[\Sigma(i\omega_n)]$ are depicted with lines.
}
\label{Fig_3}
\end{figure}

Overall, our results agree well with previous electronic structure calculations of the bulk bilayer LNO \cite{Zhang_2023a,Shilenko_2023,Lechermann_2023,
Christiansson_2023,Liao_2023,Shen_2023b,Ryee_2024,
Craco_2024,Cao_2024,YYang_2023,Qin_2023,Leonov_2026}. We note the crucial importance of strong electron-electron correlations associated with orbital-dependent quasiparticle mass renormalizations and strong incoherence of the Ni $e_g$ orbital states near $E_F$. In Fig.~\ref{Fig_3} we display the $\mathrm{DFT}+\mathrm{DMFT}$ results for the Ni $3d$ self-energies evaluated on the Matsubara axis and those analytically continued to the real energy domain using Pad\'e approximants. The Ni $3d$ self-energies obtained by $\mathrm{DFT}+\mathrm{DMFT}$ show a typical Fermi liquid-like behavior across the entire range of strains, with an orbital-dependent quasiparticle damping of $-\mathrm{Im}[\Sigma(i\omega_n)] \sim 0.23$ and 0.31~eV for the Ni $x^2-y^2$ and 
$3z^2-r^2$ states at the first Matsubara frequency, at $T = 290$ K (for the unstrained lattice). Using Pad\'e extrapolation for the self-energy to $i\omega_n \rightarrow 0$, we find about 0.11 and 0.14~eV for the Ni $x^2-y^2$ and $3z^2-r^2$ orbitals, respectively (shown in Table~\ref{tab_2}).

Our results for the quasiparticle mass renormalizations evaluated from fitting the self-energy results with Pad\'e approximants as 
$m^*/m = [1 - \partial \mathrm{Im}[\Sigma(i\omega)]/\partial i\omega]|_{i\omega\rightarrow 0}$ are given in Table~\ref{tab_2}. 
For the unstrained lattice, we obtain $m^*/m \sim 2.9$ and 3.6 for the Ni $x^2-y^2$ and $3z^2-r^2$ orbital states, implying that the Ni $e_g$ orbitals are strongly correlated. In agreement with this, the corresponding quasiparticle bands are seen to be strongly renormalized with respect to the DFT results. It is interesting that the $3z^2-r^2$ orbital states are more correlated and incoherent-like than the planar $x^2-y^2$ orbitals, in agreement with previous estimates for the bulk bilayer LNO. We explain this by a more narrow ``noninteracting'' DFT bandwidth of the $3z^2-r^2$ states, which is by $\sim$25\% smaller compared to the $x^2-y^2$ states (for the unstrained lattice). The fully occupied Ni $t_{2g}$ states are sufficiently coherent and are weakly renormalized, with $m^*/m \sim 1.3$. Overall, this underlines the importance of orbital-selective localization of the Ni $3d$ electrons in LNO.

\begin{table}[h]
{
\centering
\caption{Bonding-antibonding splitting of the Ni $3z^2-r^2$ orbitals ($\Delta_{ab}$), along with the orbitally-resolved intra-layer ($t^{||}$) and inter-layer ($t^\perp$) hopping matrix elements of the Ni ions, calculated for different strain levels.}

\begin{ruledtabular}
\begin{tabular}{lccccc}

\multicolumn{1}{c}{Strain (\%)} &  
\multicolumn{1}{c}{$\Delta_{ba}$ (eV)}  &  
\multicolumn{4}{c}{hoppings (meV)}  \\

 & & $t^\perp_{z^2/z^2}$ & $t^{||}_{x^2-y^2/x^2-y^2}$ & $t^{||}_{x^2-y^2/z^2}$ &
  $t^{||}_{z^2/z^2}$ \\

\hline
$-4$ (comp.) &  1.20  &   -599  &  -515  & -201 &  -83  \\
$-2$              &  1.18  &   -591 &  -482 &  -216  & -98   \\
 0                   &  1.16  &  -579  & -447  & -224  & -110  \\
 2                   &  1.13  &  -563  & -415 &  -225  & -120  \\
 4 (tensile)      &  1.10  &  -549  & -385 &  -222  & -128  \\
\end{tabular}
\end{ruledtabular}
\label{tab_1}}
\end{table}

\begin{table}[h]
{
\centering
\caption{Orbitally-resolved quasiparticle damping $-\mathrm{Im}[\Sigma(i\omega_n)]$ as extrapolated to $i\omega \rightarrow 0$~eV and that at the first fermionic Matsubara frequency $\omega=\pi k_\mathrm{B}T$ (given in parenthesis), and orbital-dependent quasiparticle band renormalizations $m^*/m$ of the Ni $x^2 - y^2$ and $3z^2 - r^2$ orbitals obtained within $\mathrm{DFT}+\mathrm{DMFT}$ with the Hubbard $U=6$~eV and Hund's coupling $J=0.95$~eV at $T=290$~K.}

\begin{ruledtabular}
\begin{tabular}{lcccc}

\multicolumn{1}{c}{Strain (\%)} &  \multicolumn{2}{c}{$-\mathrm{Im}[\Sigma(i\omega_n)]$ (eV)}  &  \multicolumn{2}{c}{$m^*/m$}  \\

 & $x^2-y^2$ & $3z^2-r^2$ & $x^2 - y^2$ & $3z^2 - r^2$ \\

\hline
$-4$ (comp.) &  0.02  (0.11) & 0.02  (0.15) & 2.29 & 2.73 \\
$-2$ &  0.09  (0.19) & 0.09  (0.25) & 2.53 & 3.29 \\
 0  &  0.11  (0.23) & 0.14  (0.31) & 2.92 & 3.57 \\
 2  &  0.10  (0.25) & 0.14  (0.33) & 3.30   & 3.86 \\
 4 (tensile) &  0.09  (0.26) & 0.13  (0.33) &  3.70 & 4.14 \\
\end{tabular}
\end{ruledtabular}
\label{tab_2}}
\end{table}

Our results for the unstrained lattice show that the Ni $e_g$ orbitals are nearly half-filled, with the Ni $x^2-y^2$ and $3z^2-r^2$ orbital occupations of $\sim$0.55 and 0.59 per spin orbital, respectively. The calculated total Ni $3d$ Wannier orbital occupation is about 8.25, implying a nearly 2+ oxidation state of nickel which differs from the nominal Ni$^{2.5+}$ valence state in the undoped LNO. In agreement with this, our analysis of the weights of the local Ni $3d$ orbital multiplets shows that the electronic state of the Ni ions is governed predominantly by the Ni $3d^8$ configuration (its weight is about 0.56), with an additional contribution from the $3d^9\underline{L}$ state with an oxygen ligand hole ($\sim$0.33).
This highlights the importance of the effects of small (or negative) charge-transfer energy, in agreement with strong hybridization between the Ni $3d$ and O $2p$ orbitals. Moreover, we note the emergence of a flat-band behavior (a van Hove anomaly in the {\bf k}-resolved spectral function) associated with the Ni $x^2-y^2$ states near the Brillouin zone (BZ) X-point 
close to the Fermi level. Interestingly, the Ni $x^2-y^2$ flat-band states remain essentially unaffected by strain, though they exhibit enhanced coherence under compressive strain (see Fig.~\ref{Fig_1}). We note that in comparison with the DFT band structure results, the effects of electron-electron correlations lead to a significant shift of the Ni $x^2-y^2$ flat-band quasiparticle states toward $E_{F}$, driven by the orbital-dependent bandwidth narrowing.

\begin{figure}
\centerline{\includegraphics[width=0.45\textwidth,clip=true]{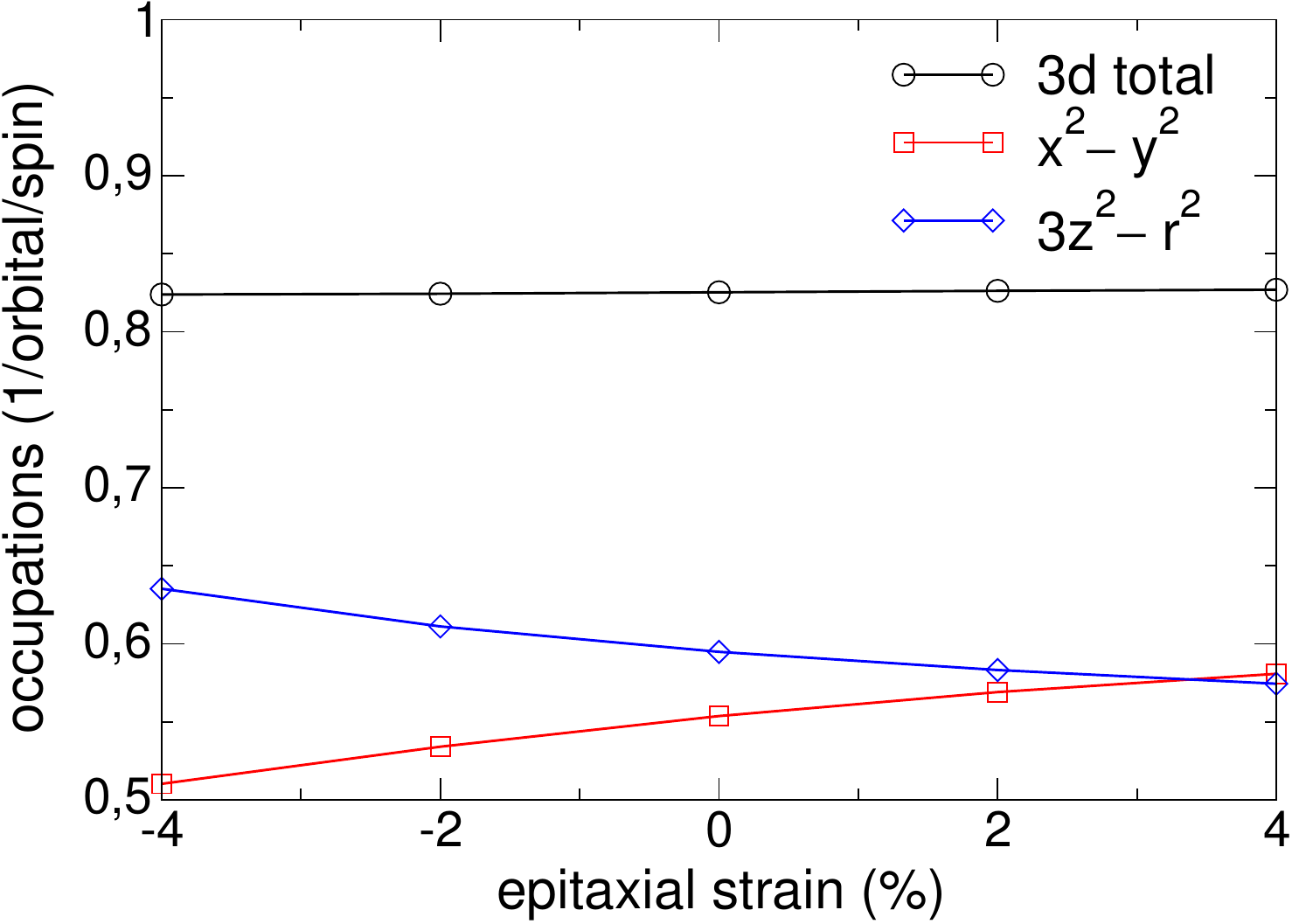}}
\caption{Our results for the total Ni $3d$ and partial Ni $x^2-y^2$ and $3z^2-r^2$ orbital occupations (per spin orbital) for different in-plane strain levels, calculated by $\mathrm{DFT}+\mathrm{DMFT}$ for PM LNO at $T = 290$~K.}
\label{Fig_4}
\end{figure}

In agreement with previous studies, our $\mathrm{DFT}+\mathrm{DMFT}$ results exhibit a continuous shift toward lower binding energies of the Ni $3z^2-r^2$ orbital states upon compressive strain (see Fig.~\ref{Fig_1})  \cite{L.Bleys_2025}. 
Under compressive strain, we find a remarkable incresease in the bonding-antibonding splitting of the Ni $3z^2-r^2$ orbitals $\Delta_{ab}$, accompanied by a sizable enhancement of the corresponding inter-layer hoppings ($t^\perp_{z^2/z^2}$), as shown in Table~\ref{tab_1}.  However, the main effect is associated with a substantial enchancement by $\sim$11\% of the intra-layer  hopping matrix elements between the planar $x^2-y^2$ orbitals, $t^{||}_{x^2-y^2/x^2-y^2}$ from $-447$ to $-515$~meV. This implies a sizable increase of the Ni $x^2-y^2$ bandwidth and the corresponding superexchange couplings.
This behavior is accompanied by a significant increase of the Ni $x^2-y^2$ and $3z^2-r^2$ orbital crystal field splitting caused by the lattice distortions (strain). It leads to a remarkable redistribution of the charge density between the Ni $x^2-y^2$ 
and $3z^2-r^2$ orbitals. In fact, the Ni $e_g$ orbital occupations exhibit a strong sensitivity to the lattice strain (see Fig.~\ref{Fig_4}). Under a compressive strain of $-4$\% the Ni $3z^2-r^2$ orbital occupations increase by about 6.8\%, to 0.64, in comparison to the undistorted case. The corresponding Ni $x^2-y^2$ orbital occupation is $\sim$0.51. That is, the orbital polarization between the Ni $x^2-y^2$ and $3z^2-r^2$ states is of about $\sim$0.1. In contrast, a tensile strain yields a nearly equally populated Ni $x^2-y^2$ and $3z^2-r^2$ orbital states, 0.58 and 0.57, respectively, with a negligible orbital polarization (for a tensile strain of 4\%). We find that the total Ni $3d$ Wannier orbital occupations are nearly unaffected by the strain and are about 8.25 per spin orbit across the entire range of strains.

\begin{figure}
\centerline{\includegraphics[width=0.5\textwidth,clip=true]{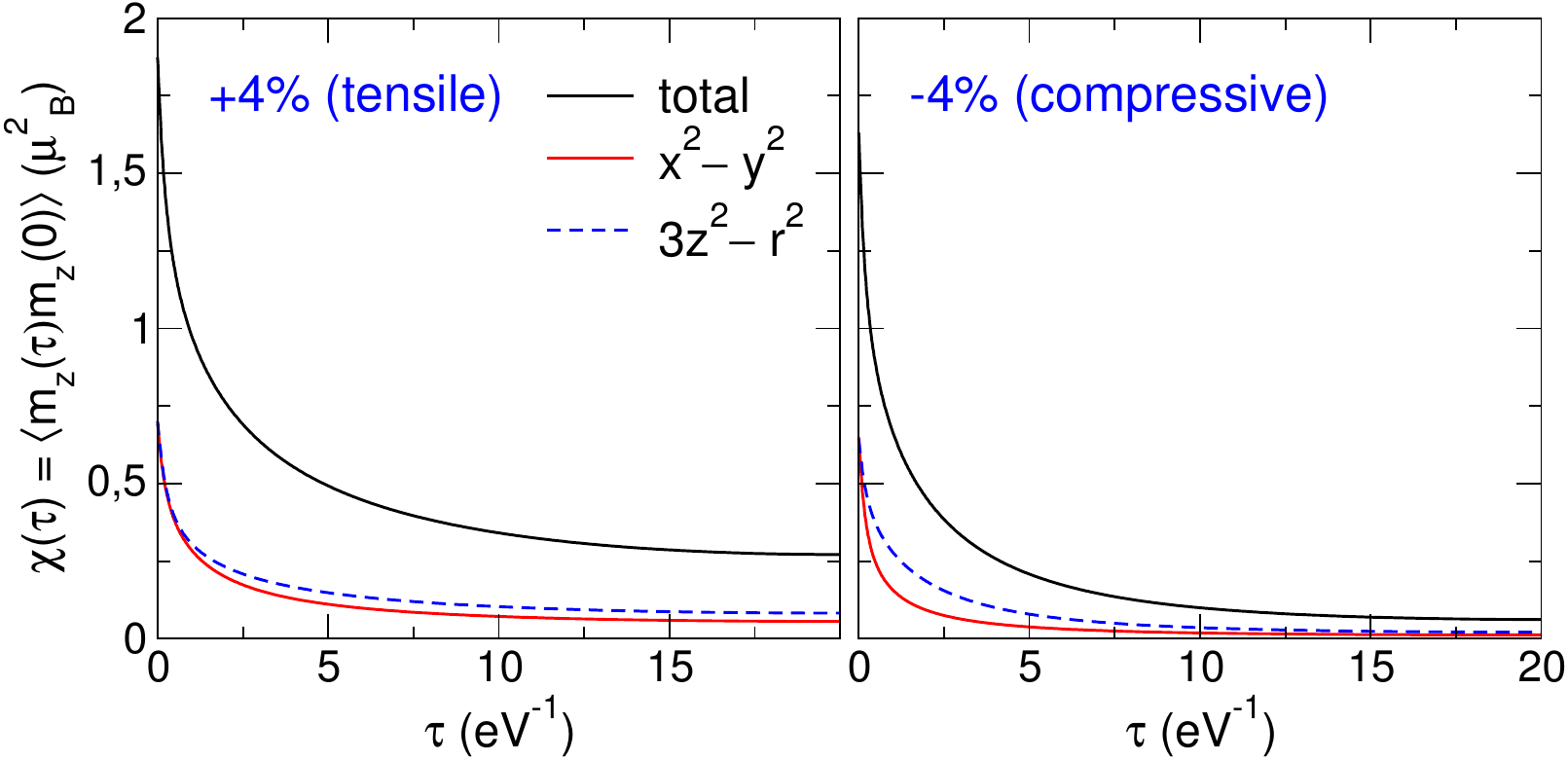}}
\caption{
Orbitally-resolved local spin correlation function $\chi(\tau) = \langle \hat{m}_z(\tau) \hat{m}_z(0) \rangle$ for the Ni $3d$ orbitals calculated by $\mathrm{DFT}+\mathrm{DMFT}$ for PM LNO under a tensile strain of 4\%  (left panel) and a compressive strain of $-4$\%  (right) at $T = 290$~K.}
\label{Fig_5}
\end{figure}

Most interestingly, the orbital-dependent quasiparticle band renormalizations $m^*/m$ exhibit a high sensitivity to the lattice strain. Under applying a tensile strain of 4\%, the Ni $x^2-y^2$ and $3z^2-r^2$ orbital states are seen to be strongly renormalized, $m^*/m \sim 3.7$ and 4.14, respectively. The Ni $x^2-y^2$ and $3z^2-r^2$ orbital states exhibit a pronounced incoherence of the spectral weights, with a large quasiparticle damping $-\mathrm{Im}[\Sigma(i\omega_n)]$ of about 0.26 and 0.33 eV at the first Matsubara frequency, respectively (at $T = 290$ K). Note that the planar Ni $x^2-y^2$ orbitals are less correlated and more coherent than the $3z^2-r^2$ orbitals. By contrast, a compressive strain of the lattice results in a significant reduction of $m^*/m$. The Ni $x^2-y^2$ and $3z^2-r^2$ quasiparticle renormalizations under compressive strain by $-4$\% are nearly the same, $\sim$2.3 and 2.7, respectively. Interestingly, this implies a decay of orbital-selective correlations, despite the significantly different occupations of the Ni $x^2-y^2$ and $3z^2-r^2$ orbitals. This is accompanied by a pronounced enhancement of coherence of the Ni $e_g$ states, associated with band-like behavior of the low-energy electronic states. In agreement with this we observe a large reduction in the strength of electron-electron 
correlations in both the Ni $x^2-y^2$ and $3z^2-r^2$ orbitals under compressive strain. 

This result corroborates with our analysis of the orbital-dependent local spin susceptibility $\chi(\tau)$ on the imaginary time domain $\tau$, evaluated within $\mathrm{DFT}+\mathrm{DMFT}$. We note that for a tensile strain of 4\% the Ni $3d$ states are seen to be close to localization with slow decaying $\chi(\tau)$ from 1.82 $\mu_\mathrm{B}^2$ to 0.27 $\mu_\mathrm{B}^2$ at $\tau=\beta/2$. This gives a relatively large value of the fluctuating moment of about 0.67$\mu_\mathrm{B}$, evaluated as $M_\mathrm{loc} = [k_BT \int \chi(\tau)d\tau]^{1/2}$. Here, $\chi(\tau)=\langle \hat{m}_z(\tau) \hat{m}_z(0)\rangle$ is the local spin-spin correlation function. The corresponding instantaneous magnetic moment $\sqrt{ \langle \hat{m}_z^2\rangle}$ is about 1.35$\mu_\mathrm{B}$. By contrast, under a compressive strain of $-4$\%, the Ni $x^2-y^2$ and $3z^2-r^2$ orbital states are seen to be less localized, i.e., more band-like. The calculated fluctuating and instantaneous magnetic moments are 0.44 and 1.28$\mu_\mathrm{B}$, respectively. The fully occupied Ni $t_{2g}$ orbitals show a nearly strain-independent band renormalization factor $m^*/m$ of $\sim$1.34 and coherent, band-like behavior. In addition, we note that the Ni $e_g$ quasiparticle band renormalizations show a sensetive depence on the choice of the Hubbard $U$ and Hund's $J$ parameters. For $U=4$~eV and $J=0.95$~eV, our $\mathrm{DFT}+\mathrm{DMFT}$ calculations yield values $\sim$1.77 and 2.23 for the Ni $x^2-y^2$ and $3z^2-r^2$ orbital states, respectively. In contrast, for $U=4$~eV and $J=0.45$~eV we obtain significantly weaker band renormalizations of $\sim$1.38 and 1.56.

\begin{figure}
\centerline{\includegraphics[width=0.5\textwidth,clip=true]{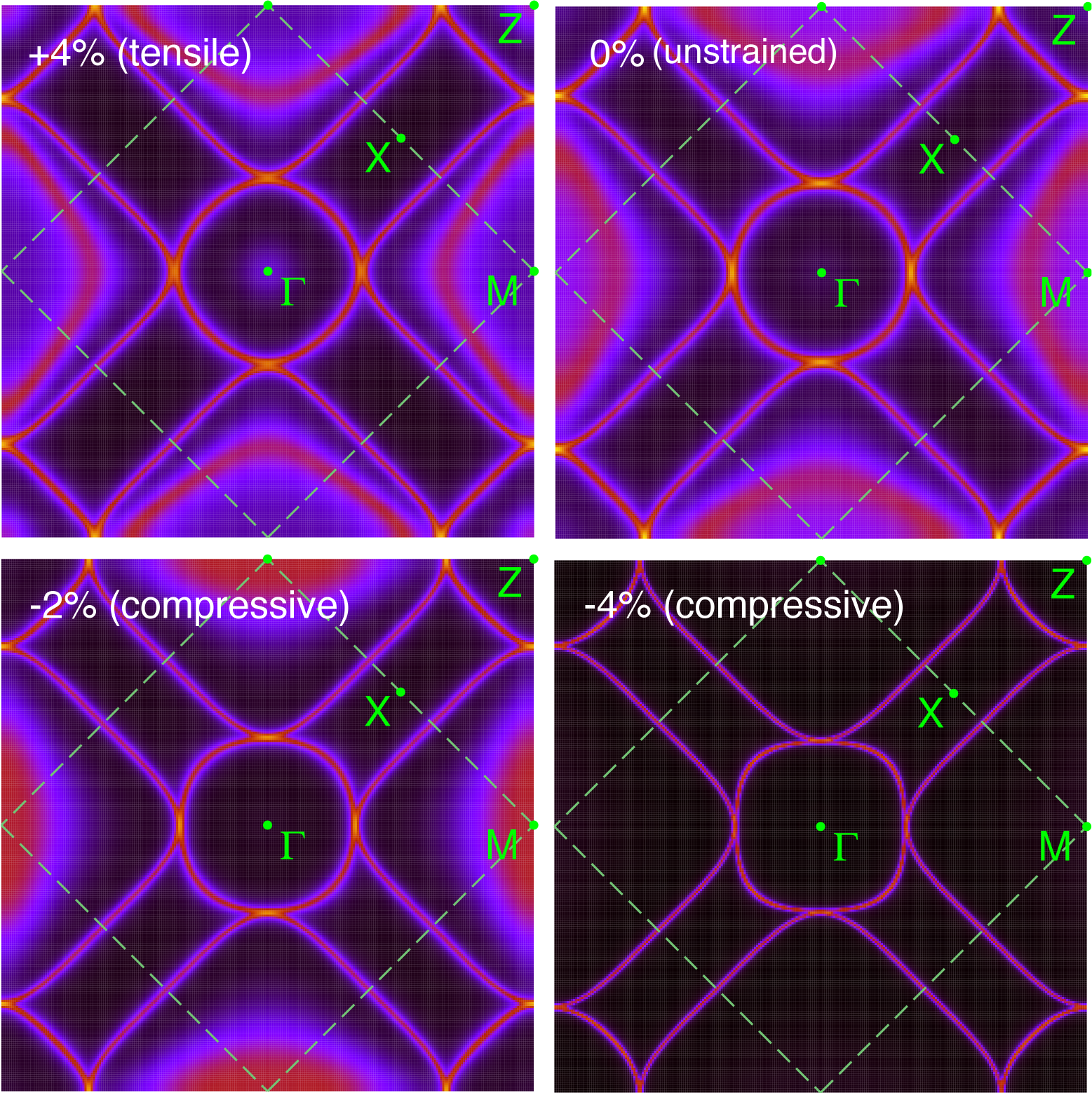}}
\caption{
Correlated Fermi surfaces of PM LNO upon applying in-plane lattice strain calculated by $\mathrm{DFT}+\mathrm{DMFT}$ at $T = 290$~K.
}
\label{Fig_6}
\end{figure}

In Fig.~\ref{Fig_6} we show our results for the quasiparticle Fermi surfaces of PM LNO obtained within $\mathrm{DFT}+\mathrm{DMFT}$ at $T=290$~K for different strain levels. The FSs are evaluated as the {\bf k}-resolved spectral function $A({\bf k},\omega)$ at $\omega=0$ for the analytically continued on the real energy axis $\omega$ self-energies $\Sigma(\omega)$. The calculated FSs are two-dimensional, closely resemble previously 
reported results for the bulk bilayer LNO \cite{Zhang_2023a,Shilenko_2023,Lechermann_2023,
Christiansson_2023,Liao_2023,Shen_2023b,Ryee_2024,
Craco_2024,Cao_2024,YYang_2023,Qin_2023,Leonov_2026}. We find that for a broad range of the in-plain strain from 4\% (tensile) to $-2$\% (compressive) both the Ni $x^2-y^2$ and $3z^2-r^2$ orbitals are partially occupied, resulting in the presence of the so-called $\gamma$ Fermi surface sheet near the BZ corner M-point. The latter is associated with a nearly fully occupied, shallow flat-band of the bonding Ni $3z^2-r^2$ orbital character. Upon further increase of compressive strain to $-4$\%, the Ni $3z^2-r^2$ quasiparticle band near the BZ M-point shifts below the $E_F$, resulting in a Lifshitz transition (associated with the suppression of the $\gamma$ FS sheet). As a result, the Ni $x^2-y^2$ and $3z^2-r^2$ orbital states yield an emergent flat-band behavior just below the Fermi level, as seen in Fig.~\ref{Fig_1}. Our results show a remarkable enhancement of the coherence of the calculated FSs, associated with the suppression of electronic correlations under compressive strain. Indeed, this result agrees with a significant decrease of $m^*/m$ for the Ni $e_g$ orbitals.

\begin{figure}
\centerline{\includegraphics[width=0.5\textwidth,clip=true]{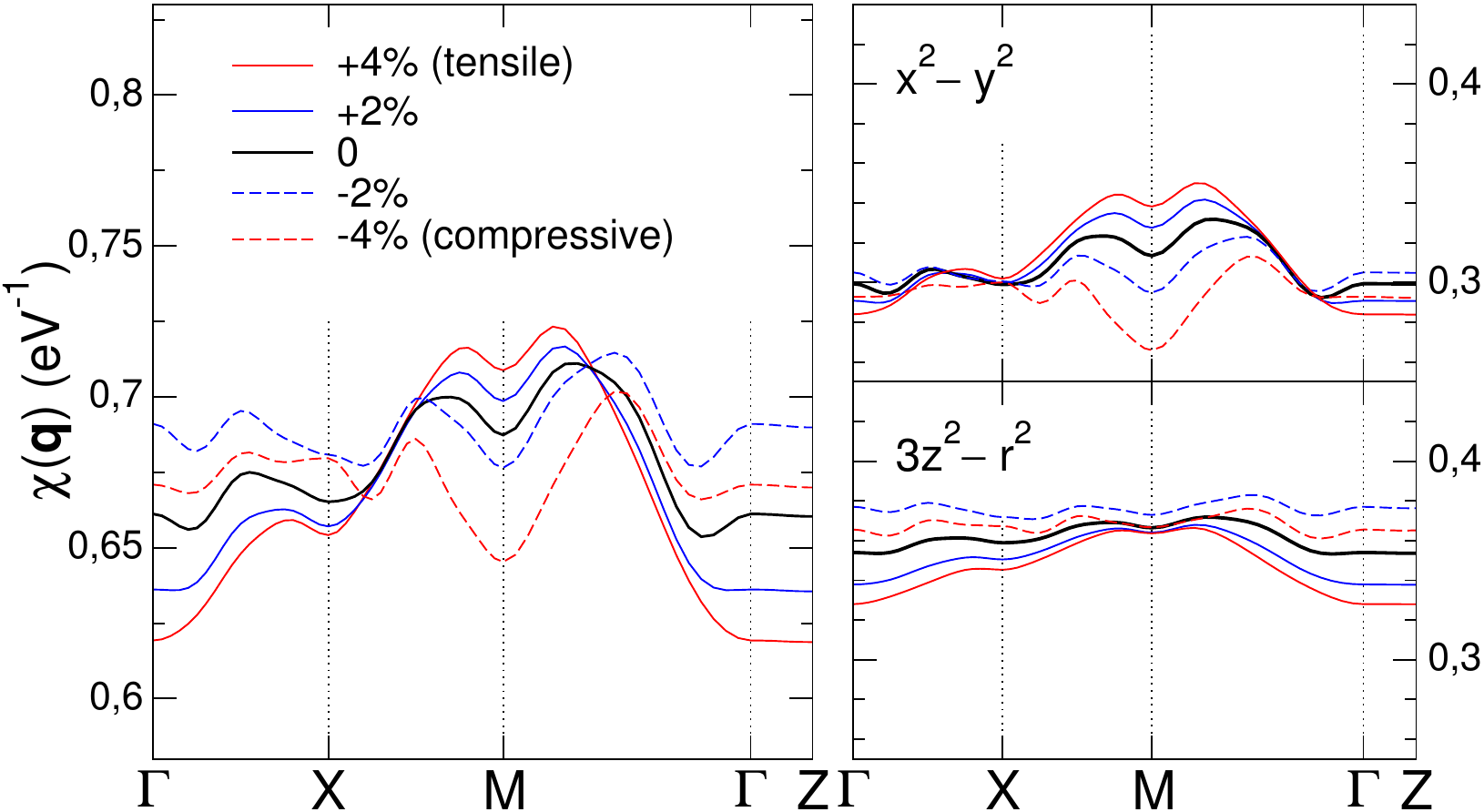}}
\caption{
Orbitally resolved in-plane static spin susceptibility $\chi({\bf q)}$ for PM LNO upon applying in-plane lattice strain at $T = 290$~K. Our calculations are performed on the Matsubara axis using $\mathrm{DFT}+\mathrm{DMFT}$ at $T = 298$~K within the particle-hole bubble approximation, neglecting the high-order vertex corrections.
}
\label{Fig_7}
\end{figure}

The $\mathrm{DFT}+\mathrm{DMFT}$ FSs exhibit multiple in-plane nesting behaviors, suggesting the possible emergence of competing long-range density wave orderings at low temperature. 
In Fig.~\ref{Fig_7} we show the momentum-resolved static magnetic susceptibility $\chi({\bf q})$ for different strain levels, evaluated using the particle-hole bubble approximation within $\mathrm{DFT}+\mathrm{DMFT}$ (neglecting high-order vertex corrections) as $\chi({\bf q}) = -k_BT \mathrm{Tr} \Sigma_{{\bf k}, i\omega_n}G_{\bf k}(i\omega_n)G_{{\bf k}+{\bf q}}(i\omega_n)e^{i\omega_n0^+}$. Here, $G_{\bf k}(i\omega_n)$ is the local interacting Green's function for the Ni $3d$ states computed on the Matsubara frequency domain $i\omega_n$ for a given momentum {\bf k}. $\chi({\bf q})$ shows complex structures with multiple well-defined maxima. This suggests a potential instability of the PM metallic state of LNO toward a long-range (intertwined) spin- and charge-density-wave state. We note that magnetic correlations are associated with incommensurate wave vectors near the BZ M-point and on the BZ $\Gamma$-X branch. The most pronounced instability is associated with a momentum {\bf k} located at the $\Gamma$-M branch, which strongly competes with a minor instability at the X-M. Our results for $\chi({\bf q})$ are in overall agreement with those previously obtained for the bilayer and trilayer nickelates \cite{Shilenko_2023,Leonov_2026,Leonov_2024a}. 
It was shown for the low-pressure bilayer LNO that similar behavior is associated with the emergence of a double spin-charge-density-wave stripe ordering characterized by a propagating wave vector $\mathrm{\bf q}=(\frac{1}{4},\frac{1}{4})$ arrangement oriented at 45$^\circ$ to the Ni-O bond diagonal spin-charge stripes \cite{Chen_2024a,Kakoi_2024,Agrestini_2024,Dan_2024,Leonov_2025,LaBollita_2024b,BZhang_2024,Ni_2024,Tian_2025}. 

Moreover, we find a heightened sensitivity of magnetic correlations to a lattice strain. We note that both a tensile and a moderate compressive strain up to about $-2$\% result in a significant enhancement of the strength of magnetic correlations from that for the unstrained LNO. Moreover, we observe that the major instability peak of $\chi({\bf q})$ shifts toward the BZ M point under tensile strain, whereas under a compressive strain of $-2$\% it appears near to $(\frac{1}{4},\frac{1}{4})$, in close similarity to the bulk bilayer LNO \cite{Shilenko_2023,Leonov_2026,Leonov_2024a}. As a result, we expect a sharp increase of planar spin- and charge-density-wave fluctuations in the strained LNO. Upon a further increase of a compressive strain to $-4$\%, we find a sharp decrease of $\chi({\bf q})$, which is apparently associated with the Lifshitz transition. It is associated with a suppression of $\chi({\bf q})$ for the Ni $x^2-y^2$ orbital states (see Fig.~\ref{Fig_7}). This suggests that both the tensile and compressive strains should, in principle, support a higher critical temperature $T_c$. However, a strong enhancement of localization for the Ni $e_g$ orbital states under the tensile strain should, in principle, promote a long-range magnetic ordering, which acts against superconductivity, resulting in the supression of $T_c$. Under a large compressive strain of about $-4$\%, LNO undergoes a reconstruction of the low-energy electronic structure, associated with the Lifshitz transition (with the disappearance of the $\gamma$ FS sheet). It leads to a remarkable reduction of the strength of magnetic correlations and, hence, most probably results in the suppression of superconductivity. Our results therefore support the picture of spin- and change-density-wave instability driven by the FS nesting in LNO. By applying external pressure or strain we effectively suppress long-range spin-charge-density-wave ordering, giving rise to enhanced spin fluctuations. This implies that in-plane spin fluctuations are a key driver of superconductivity in LNO. 

\section{Conclusion}

In conclusion, using the $\mathrm{DFT}+\mathrm{DMFT}$ method we study the normal-state electronic properties of the epitaxially strained LNO. We observe a remarkable orbital-selective
renormalization of the Ni $3d$ bands, which is strongly affected by the in-plain strain effects. This result is inline with our analysis of the {\bf k}-dependent spectral functions which show a significant orbital-dependent narrowing of the Ni $e_g$ quasiparticle bands and strong incoherence of their spectral weights under the in-plain strain across from 4\% to $-2$\%. Our results point to the proximity of the Ni $3d$ states to orbital-dependent localization. We note that both the tensile and the moderate compressive strain up to about $-2$\% lead to significant enhancement of magnetic correlations compared to the unstrained LNO. In principle, this suggests that both the tensile and compressive strain should result in a higher critical temperature $T_c$. However, strong localization for the Ni $e_g$ orbital states under the tensile strain is expected to promote a long-range magnetic ordering, which, in turn, results in the suppression of superconductivity.  Under further compressive strain to about $-4$\%, we observe the Lifshitz transition accompanied with the disappearance of the so-called $\gamma$ FS sheet (which is associated with a nearly fully occupied, shallow flat-band of the bonding Ni $3z^2-r^2$ orbital character). This results in a strong reduction of magnetic correlations and, hence, it is expected to lead to suppression of superconductivity. Overall, our results support the picture of spin- and change-density-wave instability driven by the FS nesting in LNO. It turns out that both pressure and strain can effectively tune (suppress or support) spin-change-density-wave ordering, giving rise to enhanced spin fluctuations. 

\section{ACKNOWLEDGMENTS}
The DFT electronic structure calculations were supported within the framework of the state assignment of the Ministry of Science and Higher Education of the Russian Federation for the IMP UB RAS. The $\mathrm{DFT}+\mathrm{DMFT}$ calculations, theoretical analysis of the electronic structure and magnetic properties were supported by the Russian Science Foundation (Project No. 25-12-00416) \cite{RSF}.

\end{document}